\documentstyle[12pt]{article}
\newcounter{eqnn} \newcounter{eqs} \newcounter{secn}
\newcounter{subn}

\def\eq{\addtocounter{eqnn}{1}\;\;\;(\theeqnn)} 
\def\lae{\;^{<}_{\sim} \;} 
\def\gae{\; ^{>}_{\sim} \;}

\def\lbar{\overline{\lambda}}
\def\tte{\tilde{t}}

\begin{document}

\begin{titlepage}

\pagestyle{empty} 
\begin{flushright}  
HELSINKI\\              July 1997\end{flushright}              
\vfill              \begin{center} {\LARGE 
Comment on Vacuum Stability and
Electroweak Baryogenesis in the MSSM with Light Stops. 
\\ } 
\end{center} \vfill \begin{center} {\bf 
J.McDonald 
\begin{footnote}{e-mail: mcdonald@phcu.helsinki.fi}\end{footnote}
}\\ 
	     \vspace {0.1in}     
Dept. of
Physics,\\P.O.Box 9,\\University of Helsinki,\\ FIN-00014
Helsinki,\\FINLAND \\ \vfill \end{center} 
\newpage \begin{center}
{\bf Abstract} \end{center}  
		
	       We show that, for all values of $Tan \beta$ and the light 
right-handed stop mass 
for which the electroweak 
phase transition is strong enough to avoid washout following
 electroweak baryogenesis,
 the electroweak vacuum is stable over the lifetime of
 the observed Universe. 
 Cosmological stability of the electroweak vacuum is violated only if 
the light right-handed stop is lighter than 100-115GeV.
\end{titlepage}

		     The possibility of producing 
the baryon asymmetry of the Universe 
at the electroweak phase transition \cite{ewb,shap} 
in the minimal supersymmetric standard model (MSSM) 
 has been the subject
 of much study in recent times \cite{emssm,cpv,ptw}. 
It has become clear that although 
the MSSM can readily provide the CP violation necessary
for baryon asymmetry generation \cite{cpv}, 
it has more difficulty producing a sufficiently 
strong first-order phase transition to 
prevent subsequent wash-out of the asymmetry 
\cite{ptw}. One possibility for having a 
Higgs mass consistent with electroweak 
baryogenesis is associated with the rather special range of
 parameters for which the soft SUSY 
breaking mass squared parameter of the mostly right-handed (r.h.) 
stop partially cancels 
against the finite-temperature contribution 
to its mass, in order that the contribution from the Higgs expectation value 
dominates the r.h. stop mass during the electroweak phase transition 
\cite{eqz,cwe}. (It has also been recently 
suggested that the inclusion of two-loop QCD thermal corrections to the 
effective potential of the Higgs field 
may sufficiently increase the strength of the electroweak phase transition to 
evade wash-out, even with a 
positive mass squared for the r.h. stops \cite{ebc, ml2}). 
Since this requires a negative soft SUSY breaking 
mass squared for the r.h. stop, 
there will typically be a second minimum of the effective potential
with a non-zero stop expectation value. 
For the values of the parameters of the model which give the largest 
possible Higgs mass consistent with electroweak baryogenesis,
 this vacuum is generally of lower energy than the electroweak vacuum. 
In this case it is important to
consider the cosmological stability of the 
electroweak vacuum i.e. whether it is longer lived than the 
observed Universe. In a previous 
analysis \cite{cwe} it was shown that the upper bound obtained 
on the Higgs mass in the MSSM is strongly dependent on whether
 or not it is possible to have a 
sufficiently long-lived metastable electroweak vacuum.
If it is not possible, then the upper bound obtained 
from the 1-loop resummed finite-temperature 
effective potential is around 80GeV \cite{cwe}.
 On the other hand, 
if we can have a 
metastable vacuum over the whole range of 
parameters for which the electroweak phase 
transition is consistent with electroweak 
baryogenesis, then the Higgs mass could be as large as 100GeV. 
The actual lifetime of the 
metastable electroweak vacuum was not, however, calculated. 
In this short letter we will consider the 
lifetime of the metastable electroweak 
vacuum and show that it is indeed cosmologically 
stable over the range of parameters for which 
electroweak baryogenesis is possible.

	We consider the effective theory 
consisting of the Standard Model (SM) fields
 plus light r.h. stops. All other 
fields will be considered to be too heavy to contribute to 
the finite-temperature effective potential. 
This corresponds to the limit of large pseudo-scalar Higgs mass $ m_{A}$ and 
heavy left-handed (l.h.) stop with negligible left-right stop mixing,
 which is known to give the electroweak 
baryogenesis upper bound on the Higgs mass 
in the MSSM in this case \cite{cwe}. 
The effect of supersymmetry is then to 
fix the lightest Higgs mass in terms of 
$Tan \beta$. At 1-loop the Higgs mass is 
given by \cite{cwe}
$${\rm m_{H}^2 = m_{Z}^{2} Cos^{2} 2\beta + \frac{3 m_{t}^4}{4 \pi^2 v^2} 
ln \left( \frac{m_{\tte}^2 m_{\tilde{T}}^2}{m_{t}^4} \right)   \eq},$$
where $m_{\tilde{T}}$ is the mass of the 
heavy l.h. stop, $m_{t}$ is the top quark mass, 
$m_{\tte}$ is the mass of the light r.h. stop 
($ m_{\tte} = (-m_{u}^{2}+\lambda_{t}^{2} v^{2}/2)^{1/2}$), 
$\lambda_{t} \approx 1$ is the SM top quark Yukawa coupling
and $v$ is the 
Higgs vacuum expectation value ($v = 250GeV$).
The potential of the Higgs 
($\phi$) and r.h. stop ($U$) field in this effective theory is given by 
$${V(\phi, U) = -\frac{m_{\phi}^{2}}{2} \phi^2 + \frac{\lambda}{4} \phi^4 
- \frac{m_{U}^2}{2} U^2 + \frac{\lambda_{t}}{4} \phi^2 U^2
 + \frac{g_{3}^{2}}{24} U^4     \eq}.$$
(We use canonically normalized real scalar fields throughout).
This potential has minima at $(\phi_{o},0)$ and $(0,U_{o})$,
 seperated by a "ridge" which has a 
minimum height at a saddle point given by $(\phi_{m},U_{m})$, where \cite{cwe}
$${ \phi_{o}^2 = \frac{ m_{\phi}^2}{\lambda}    \eq},$$
$${ U_{o}^2 = \frac{6 m_{U}^2}{g_{3}^2}    \eq},$$
$${\phi_{m}^2 = \frac{2 (m_{\phi}^2 - \frac{3 m_{U}^2 
\lambda_{t}^2}{g_{3}^2})}{2 \lambda - 
\frac{3 \lambda_{t}^4}{g_{3}^2}}    \eq}$$
and
$${U_{m}^2 = \frac{2 ( m_{u}^2 - \frac{m_{\phi}^2 
\lambda_{t}^2}{2 \lambda} )}{\frac{g_{3}^2}{3}
 - \frac{\lambda_{t}^4}{\lambda}}    \eq}.$$
In order to discuss vacuum stability in this model we would generally
 have to consider vacuum tunnelling with two real scalar fields. 
However, this is a difficult problem in general, since the 
simple bounce action approach \cite{cc,ll} cannot be applied here \cite{kls}. 
A recent analysis of 
charge and colour breaking minima in the MSSM approached the problem 
via a lattice minimization of a modified Euclidean action \cite{kls}.
In the present paper we 
will adopt a more direct approach. 
We will first derive upper and lower $bounds$ on the decay rate 
using the one scalar field bounce 
action approach and then argue that, so long as the 
constraints imposed on the 
model parameters by these upper and lower 
bounds are close, we can safely use them to determine 
whether or not the electroweak vacuum is cosmologically stable. 

	An upper bound may be found by simply considering the one-dimensional 
"straight line" potential 
connecting the electroweak minimum and the stop minimum
and then calculating the bounce action for this potential; 
this amounts to an Ansatz for the minimum action solution, 
obtained by fixing the 
value of the orthogonal component of the scalar field 
 and minimizing the remaining action. 
Since this will not be a true solution of the 
Euclidean equations of motion, the resulting
tunnelling action will be higher than that 
of the true solution and so we will obtain a lower bound on the vacuum decay rate.
In order to obtain an upper bound on the vacuum decay rate, we will adopt 
the following 
 proceedure. In general, the barrier at the saddle 
point will not be on the straight line connecting the minima 
and will be lower than the barrier of the straight line potential. 
Thus if we were to consider a second 
straight line potential, with the same distance in field space between the 
electroweak and 
stop 
minima and with the same energy
splitting between the vacua as for the true straight line potential 
 but with a barrier height equal to that of the saddle point, then we 
would expect the bounce action for this 
potential to be smaller than the true bounce action 
(which would have at least as large a barrier and a longer minimum action path).
 Thus by considering the bounce action 
along these two straight line potentials 
we will obtain upper and lower bounds on
the vacuum decay rate. If these upper and lower bounds turn out to be close, 
then this proceedure will give us a simple and accurate
method for estimating the vacuum decay rate. 

	       The true straight line potential is obtained by first defining 
$\phi^{'} = \phi_{o} - \phi$
and then defining a real scalar 
field $\rho$ such that $ \phi^{'} = \rho Cos \theta$ and 
$U = \rho Sin \theta$, where $Tan \theta = U_{o}/\phi_{o}$. 
The straight-line potential (which we denote by 
the subscript 1) is then given by 
$${  V_{1} = \frac{\alpha_{1}}{2}  \rho^2 - \frac{\beta_{1}}{2 \sqrt{2}}
 \rho^3 + \frac{\gamma_{1}}{4}
\rho^4    \eq},$$             
where 
$${ \alpha_{1} = 3 \lambda \phi_{o}^2 c_{\theta}^2 +
 \frac{\lambda_{t}^{2}}{2} \phi_{o}^{2}s_{\theta}^2
- (m_{\phi}^2 c_{\theta}^2 + m_{u}^2 s_{\theta}^2)    \eq},$$
$${ \beta_{1} =   \sqrt{2} \phi_{o} c_{\theta} ( 2 \lambda c_{\theta}^2
 + \lambda_{t}^2 s_{\theta}^2)  \eq}$$
and
$${\gamma_{1} = \lambda c_{\theta}^4 + 
\lambda_{t}^2 
c_{\theta}^2 s_{\theta}^2 + \frac{g_{3}^2}{6} s_{\theta}^4 \eq}.$$
$\Delta V$, the splitting in the potential energy, and 
$\rm \rho_{o}$, the minimum of the potential, are given by 
$${ \Delta V = -\left(\frac{\alpha_{1}}{2} \rho_{o}^2 
- \frac{\beta_{1}}{2 \sqrt{2}}
 \rho_{o}^3 + \frac{\gamma_{1}}{4} \rho_{o}^4\right)    \eq}$$             
and
$${ \rho_{o} = \frac{1}{4 \sqrt{2} \gamma_{1}} \left(3 \beta_{1} 
+ \sqrt{9 \beta_{1}^2 - 32 \alpha_{1} \gamma_{1}} \right)   \eq}.$$
We next modify $\alpha$, $\beta$ and $\gamma$
in order to obtain a second potential, $V_{2}$, 
with the same values of $\Delta V$ and $\rho_{o}$ but
 with a maximum of the potential equal 
to the height of the saddle point on the ridge 
seperating the two minima. The saddle point barrier height is given by 
$${ V_{b} = - \frac{3 (\lambda_{t}^2 m_{\phi}^2 
- 2 \lambda m_{u}^2)^2}{4 \lambda (2 \lambda g_{3}^2 - 3 \lambda_{t}^4)}
	     \eq}.$$
In this case we find that the potential parameters are given by 
$${ \alpha_{2} = \frac{2}{\rho_{o}^2} \left( 
\frac{\gamma_{2}}{4} \rho_{o}^4 - 3 \Delta V\right) \eq},$$
$${ \beta_{2} = \frac{4 \sqrt{2}}{\rho_{o}^3} 
\left( \frac{\gamma_{2}}{4} \rho_{o}^4 -  \Delta V\right)     \eq}$$
and
$${\frac{64 V_{b}\gamma_{2}^3}{\rho_{o}^4}  
= (\gamma_{2} + y)(\gamma_{2} - 3 y)^3    \eq},$$
where
$${ y = \frac{4 \Delta V}{\rho_{o}^4}   \eq}.$$

	  We next consider the bounce action for these two potentials. 
The bounce action is given by 
$${ S_{4} = \frac{2 \alpha}{\beta^2} \tilde{S}_{4} 
(\overline{\lambda}) \eq},$$
where $\tilde{S}_{4} (\overline{\lambda})$ 
is the rescaled bounce action \cite{ll} and
 $${ \overline{\lambda} = \frac{4 \alpha}{\beta^2} \gamma \eq}$$
is the scalar self-coupling in the rescaled theory.
We find that we can accurately fit 
$\tilde{S}_{4} (\overline{\lambda})$  for $ \lbar \leq 0.8$ by
$${ \tilde{S_{4}}(\lbar) \approx \rm a_{1} + a_{2} e^{x} + a_{3} e^{2x} 
+ a_{4} x + a_{5} x e^{x} + a_{6} x e^{2x} + a_{7} x^2 +
 a_{8} x^2 e^{x}  + a_{9} x^2 e^{2 x}    \eq},$$
where the values of the $a_{i}$ are 
given in Table 1. In practice, for the cosmologically
unstable vacuum of interest we find that $\lbar \lae 0.6$.
(The field at the centre of the 
corresponding true vacuum bubble, from which the bounce solution 
may be readily obtained \cite{ll}, is given by
$ \psi_{o}(\lbar) \approx 5.78 -2.77 \lbar - 0.91 \lbar^2 -0.24 \lbar^3 $).

	  Cosmological stability of the electroweak vacuum requires that  
$\rm S_{4} \gae 400$ \cite{cs}. This constraint is then compared with the 
bounce action obtained for the case where 
(i) the electroweak phase transition is 
consistent with electroweak baryogenesis 
and (ii) we arrive in correct final vacuum state. The 
first condition requires that 
the value of the scalar field at the end of the phase transition is 
large enough to suppress the 
sphaleron rate and so prevent washout of the asymmetry  
(we refer to this as the 
"no washout" constraint) \cite{ewb,shap,ha,dle}. 
This requires that at the end of the 
electroweak phase transition we have
$${ \frac{\phi_{+}(T_{1})}{T_{1}} \gae 1   \eq},$$
where $\phi_{+}(T_{1})$ is the value of $\phi(T)$ at the  
non-zero minimum of the effective 
potential, with the phase transition essentially occuring  
as soon as the minima of the finite-temperature effective
potential become degenerate at $T_{1}$ \cite{ha}. 
The second condition requires that the finite-temperature effective potential 
along the light stop direction is 
stable at the temperature of the electroweak phase transition, in order
 that there is no phase transition 
to the squark minimum (we refer to 
this as the "no squark transition" constraint)
 \cite{cwe}.

	      The no washout, 
no squark transition and vacuum stability constraints result in upper 
	      and lower bounds 
on the r.h. stop mass $m_{\tte}$ for a given $Tan \beta$, as given in Table 2. 
In calculating these bounds we considered the now standard
 1-loop resummed finite temperature effective potential \cite{ha,dle},
 for the particular case of the 
SM plus light r.h. stops and 
for values of $Tan \beta$ corresponding to $m_{H}$ 
greater than the experimental lower bound, $m_{H}\gae 65GeV$ \cite{mh}. 
(We have fixed $m_{\tilde{T}} = 500GeV$ in equation (1) throughout). 
The no squark transition constraint imposes a lower bound on $m_{\tte}$. 
In calculating this lower bound we 
considered the 
case where the $T^{3}$ term in the finite-temperature effective potential 
along the 
stop direction includes fully 
the contribution of the squarks and Higgs, 
corresponding to the case where their finite-temperature effective masses 
are dominated by the U expectation value \cite{cwe}. 
The no washout constraint imposes an upper bound on $m_{\tte}$. The upper bound 
on the Higgs mass then corresponds to the value
 of $Tan \beta$ at which these 
upper and lower bounds come together. 
We find that this occurs at $ Tan \beta \approx 4.5$, corresponding to 
$m_{H} \approx 92GeV$. (This is in broad agreement with reference \cite{cwe}; 
they also allow for light gauginos, which we have not considered here). 
In addition, we give in Table 2 the lower bound on $m_{\tte}$ coming from 
the condition that the electroweak vacuum is an absolutely stable 
global minimum of the effective potential, 
as would be necessary if the metastable 
electroweak vacuum were not 
cosmologically stable. This imposes 
an upper bound on the Higgs mass, $ m_{H} \approx 81GeV$. 
We also give the lower bound on $m_{\tte}$ coming from the 
requirement that the electroweak
vacuum is cosmologically stable. In fact we obtain two bounds, 
coming from $V_{1}$ and $V_{2}$, 
which turn out to be very close in practice.
From Table 2 we see that the lower bound on $m_{\tte}$ coming 
from the requirement of cosmological stability of the vacuum is 
much smaller then the range permitted by the no washout and 
no squark transition conditions. 
Typically, the electroweak vacuum in the 
effective theory consisiting of the 
SM plus light r.h. stops becomes cosmologically 
unstable only for $m_{\tte}$ less than 100-115GeV, depending on the value of 
$Tan \beta$. (The 
present lower bound on the stop mass is $m_{\tte} \gae 70GeV$ \cite{brg}).
 Observation of a light, mostly r.h. 
stop of mass less than 100GeV together with a 
light Higgs of mass less than about 100GeV would therefore require 
new physics at the weak scale in order to stabilize the electroweak vacuum. 
(It is 
also possible to have a light stop 
without having a negative mass squared for the r.h. stop, but this 
would require a large mixing of left and right-handed stops).

     From this we may conclude 
that the electroweak vacuum, although typically metastable for 
the case of light r.h. stops, is 
nevertheless cosmologically stable over the whole range of 
$m_{H}$ and $m_{\tte}$ consistent with electroweak baryogenesis.
Therefore the upper bound on the Higgs mass
 should be taken to be that imposed by the no squark transition 
constraint and not that coming 
from the requirement of absolute stability of the 
electroweak vacuum. 
This should allow the light Higgs to have a mass of 90GeV or more without
conflicting with electroweak baryogenesis. 

{\bf Acknowledgements} The author 
would like to thank Kari Enqvist for several useful suggestions. 
This research was supported under the 
TMR programme by EU contract ERBFM-BICT960567.

\newpage

{\bf Table 1. Coefficients of the bounce action fit}  

\begin{center}
\begin{tabular}{|c|c||c|c|}          \hline
$a_{1}$ & $1.07645192933387x10^{10}$ & $a_{2}$ & $-1.728837410146912x10^{10}$ \\
$a_{3}$ & $6.523854899000699x10^{9}$ & $a_{4}$ & $4.292928013770597x10^{9}$ \\
$a_{5}$ & $2.250364456777935x10^{9}$ & $a_{6}$ & $-2.302628340489138x10^{9}$ \\
$a_{7}$ & $5.233330738163529x10^{8}$ & $a_{8}$ & $-2.806865631829403x10^{9}$ \\
$a_{9}$ & $2.349084983026382x10^{8}$ & &    \\ \hline
\end{tabular}
\end{center} 
\vspace{15mm}
{\bf Table 2. Upper and Lower Bounds on $\bf m_{\tte}$ as a function of 
$\bf Tan \beta$. 
\newline \rm The upper bound is from the no washout constraint. 
The lower bounds are from the no squark transition constraint, from the 
condition of absolute electroweak vacuum stability ("stability") 
and from the condition 
for the electroweak vacuum to be cosmologically stable as calculated using 
$V_{1}$ and $V_{2}$ respectively. The 1-loop 
values of $m_{H}$ as a function of 
$Tan \beta$ are also given, using the 
no washout value of $m_{H}$. All masses are in GeV.}
  
\begin{center}
\begin{tabular}{|c|c||c||c|c|c|c|}    \hline
$Tan \beta$ & $m_{H}$ & $m_{\tte}$ & $m_{\tte}$ 
& $m_{\tte}$ & $m_{\tte}$ & $m_{\tte}$ \\ 
& & $\rm (no \; washout)$ & $\rm (no\; squark\; transition)$ 
& $\rm (stability)$ & $(V_{1})$  & $(V_{2})$ \\ \hline
1.7 & 65.4   & 172.9 & 155.1 & 158.1 & 113.1 & 116.3 \\  
2.6 & 81.0   & 153.3 & 147.0 & 153.3 & 107.4  & 108.1 \\  
3.0 & 84.9   & 148.2 & 145.0 & 151.8 & 105.8  & 106.2 \\  
4.5 & 92.4   & 139.8 & 139.8 & 149.5 & 102.7  & 102.7 \\ 
10.0 & 97.7   & 132.8 & 136.1 & 148.2 & 100.3  & 100.3 \\ \hline  
\end{tabular} 
\end{center}

\end{document}